# Multiple Sclerosis and Geomagnetic Disturbances: Investigating a Potentially Important Environmental Risk Factor

**Seyed Aidin Sajedi**

Neuroscience Research Center, Golestan University of Medical Sciences, Gorgan, Iran

Department of Neurology, Sayad Shirazi Hospital, Golestan University of Medical Sciences, Gorgan, Iran

Multiple Sclerosis Center, Golestan Hospital, Ahvaz, Iran

**Fahimeh Abdollahi**

Department of Internal Medicine, Golestan University of Medical Sciences, Gorgan, Iran

Corresponding author: Seyed Aidin Sajedi

Address for correspondence: Department of Neurology, Sayad Shirazi Hospital, Golestan University of Medical Sciences, Gorgan, Iran

Email: dr.sajedy@gmail.com

**Abstract**

Multiple sclerosis (MS) is the most common disabling neurological disorder among young adults. In spite of many types of research, the cause of the disease remains unclear, but a significant body of evidence indicates that environmental risk factors play key roles in this disease's etiology. Various hypotheses have been posited up to now on the presumed disease risk factors; however, they have not been successful in explaining all MS features. The aim of this article is to introduce the concept of the newly proposed geomagnetic disturbance hypothesis of MS and its ability to explain certain features of the disease in order to encourage medical geologists to contribute to this field of research.

**Background**

Multiple sclerosis (MS) is the most common disabling neurological disorder among young adults (Wallin 2008). It is regarded as an immune-mediated disorder and its manifestations are due to frequent attacks of the auto-reactive cellular immune system on components of the central nervous system (CNS), including any part of the brain or spinal cord.

MS has a heterogeneous pathophysiology. Major roles in its pathophysiology are played by auto-reactive immune T cells and the temporary dysfunction of the blood-brain-barrier (BBB), which allows these activated T cells to enter the CNS. Axons and glial cells are also impacted during attacks; however, the myelin sheath (i.e. the cover layer of axons) is the main target of the immune attack. Myelin is very important for the maintenance of proper and rapid impulse conduction through the axons. MS neurological deficits are mainly due to demyelination (i.e. destruction of myelin), and consequently, impairment in impulse conduction.

Accordingly, the disease's clinical features are dependent on the site of demyelination and axonal damage, and may vary from sensory deficits, motor impairment, vision loss and/or ataxia to cognitive dysfunction. In most cases, the disease exhibits a relapsing-remitting course, consisting of short-term attacks of neurological deficits that recover completely or near completely. However, in a smaller group of patients, the disease progressively worsens (known as primary progressive MS). The diagnosis of MS is confirmed based on clinical signs and symptoms, as well as the existence of supportive imaging evidence from magnetic resonance imaging (MRI).

Since the time that Charcot identified MS as a disease in the 19th century, it has been the subject of numerous studies, and significant funds have been spent to determine its cause; however, the cause of the disease has remained largely not understood (Murray 2005). Genetic studies of twins have revealed that the role of genes in MS is at most 30% as a factor of susceptibility, and considering the special epidemiological features of MS, the remaining factors involve one or multiple non-genetic triggers (Wallin and Kurtzke 2008). Various hypotheses have been proposed concerning the presumed nature of those triggers or risk factors. Cold temperature and humidity, toxic elements, smoking, nutritional deficits, bacterial infections, parasitic infection, and viral infections and the molecular mimicry in immune response to them, were among numerous risk factors that were posited, but previous hypotheses have not been successful in explaining MS pathophysiology or its epidemiological features (Murray 2005; Ascherio and Munger 2007, 2007).

In this article, we review the strengths and weaknesses of our proposed geomagnetic disturbance (GMD) hypothesis of MS and compare it with the ultraviolet B-related vitamin D deficiency hypothesis. To understand various aspects of GMD hypothesis, one should be familiar with certain special characteristics of MS, as well as have basic knowledge of geomagnetic activities and space-weather. Since the main audience of this article is medical geologists, we presume that explaining the basics of GMD is not necessary; however, for other biomedical researchers who feel unfamiliar with these concepts we recommend a simplified, concise and open access description about these issues in a special section of our previous publication, which is devoted to the basics of space-weather and GMD (Sajedi and Abdollahi 2012).

**Special features of MS**

As mentioned above, MS has special epidemiological features that have been tantalizing for epidemiologists and researchers attempting to identify non-genetic risk factors. First of all, MS has a well-known and notable pattern of distribution. It is more prevalent in higher latitudes of both hemispheres and is rare in equatorial areas. This feature of MS was revealed early on from large epidemiological studies and is known as the MS north-south prevalence gradient (Murray 2005; Wallin and Kurtzke 2008). Due to the recent increase in the incidence of MS in low-latitude areas, a recent study determined that this gradient has been attenuated during the past decades (Koch-Henriksen and Sörensen 2010). In spite of this fact, a subsequent study verified that -the MS north-south prevalence gradient still exists (Sajedi and Abdollahi 2012).

A second feature of MS is the effect of migration on susceptibility to the disease. Multiple studies have clarified that the risk of MS will increase or decrease if the person migrates to high or low prevalent area before the age of 15 years, respectively (Wallin and Kurtzke 2008). The migration effect has been regarded as a clue that risk factor(s) of MS should be strongly related to the environment of residence location. More interestingly, there is epidemiological evidence that the time of birth, especially the month of birth, is a determining factor in the risk of suffering from MS in adulthood. Birth in May in high latitudes of the northern hemisphere and in November In high latitudes of the southern hemisphere was found to be a risk factor for MS (Staples, Ponsonby, and Lim 2010; Willer et al. 2005).

Another well-known but controversial feature of MS is "MS epidemics". Krutzke et al. reported four separate clusters of MS incidence epidemics in the limited population of the Faroe Islands from 1940 to 1991 (Kurtzke et al., 1993). Krutzke proposed that those epidemics were a significant sign that MS is transmitted by a viral infection. However, despite many efforts using sensitive techniques, the presumed pathogen has never been detected (Wallin and Kurtzke, 2008). Developing a hypothesis to explain these epidemics by means of a non-infectious environmental risk factor has been a significant challenge for researchers.

Another interesting feature of MS is the existence of *seasonality* in its attacks. There are multiple reports that MS attacks most often occur during spring and summer, especially in March-April and August-September (Meier et al. 2010).

Finally, MS is notable for its patterns of historical incidence. MS was regarded as a very rare disease in the 19th century. In the early decades of the 20th century, MS incidence rose rapidly. This increase was regarded by some researchers as "an epidemic of recognition rather than the effect of altered biological factors" (Murray, 2005). But from the 1930s, rising MS incidence was reported initially from the high latitudes of the northern hemisphere and then from the southern hemisphere. This trend became even more pronounced from 1945 to 1954 (Midgard et al., 1996). Subsequently, MS incidence grew continually and universally, an indicator that exposure to MS environmental risk factors has actually altered. The growth in MS incidence, especially the increase in the first half of the 20th century, has been confirmed in a recent birth cohort by Ajdacic-Gross et al. (Ajdacic-Gross et al., 2013).

**Critique of the current hypothesis of MS: Sunlight exposure and vitamin D hypothesis**

Based on the features mentioned above, it seems obvious that the environmental risk factor(s) of MS should be strongly dependent on time, place, and geophysical features. Therefore, it is reasonable that many researchers are convinced that a relationship may exist among MS and sunlight exposure.

It has been previously demonstrated that temperature is not a risk factor for MS. For this reason, in the opinion of many biomedical researchers, the only remaining factor that changes relative to latitude and time is the received solar radiation. The effect of sunlight and its related ultraviolet B-dependent production of vitamin D in the body is commonly known. Accordingly, it has been hypothesized that vitamin D deficiency due to insufficient sunlight exposure may be the cause of triggering MS in susceptible individuals. By this explanation, it would be possible to describe why MS prevalence increases by moving from the equator toward the poles.

Based on the vitamin D hypothesis (VDH), the effect of time of birth can be explained by insufficient production of vitamin D in pregnant mothers, and consequently, in the fetus during winter time. Therefore, neonates who are born with such deficiency in the spring would be at higher risk of MS in the future (Willer et al. 2005; Staples, Ponsonby, and Lim 2010). For this reason, efforts to find reasonable evidence to support the VDH, as the ultimate solution to the mystery of MS, increased significantly. In this regard, a study on experimental allergic encephalitis (EAE), which is regarded as an animal model of MS, showed that by supplementing the affected animals with vitamin D, EAE will significantly alleviate (Cantorna, Hayes, and DeLuca 1996). Moreover, various studies reported that vitamin D deficiency is more common in MS patients. Consequently, the majority of MS researchers were convinced that the actual risk factor of MS had been found.

At first glimpse, VDH seems to be successful in explaining the primary MS epidemiological features and pathophysiology. Nevertheless, there are key shortcomings with this hypothesis that have been neglected by researchers. The most important feature of MS that influenced researchers to relate it to sunlight exposure was the MS north-south prevalence gradient. If the VDH is true, this latitudinal gradient should be relatively linear and MS prevalence should increase constantly by moving from the equator toward the poles. However, the reality is that MS prevalence changes linearly, only in the southern hemisphere. In contrast, in the northern hemisphere, MS prevalence changes are notably parabolic. This phenomenon was identified from the first reports of this gradient (Kurtzke, 1980) up to the most recent study (Koch-Henriksen and Sörensen, 2010). Moreover, a recent modeling study has shown that the notion of the existence of a latitudinal gradient of vitamin D in the population of a particular geographical area is not necessarily accurate (Kimlin, Olds, and Moore, 2007).

More importantly, a recent study by Ueda et al. in Sweden clarified that the concept of explaining the effect of month of birth by means of VDH is flawed (Ueda et al. 2014). Because the blood samples of all Swedish neonates have been saved since 1975, researchers were able to evaluate blood vitamin D levels at the time of birth of 459 MS patients and compared them with the same sample of healthy individuals. Simultaneously, they examined their month of birth, diet, sunlight exposure and socioeconomic state. The result demonstrated that there was no relationship between the blood level of vitamin D just after birth and the risk of developing MS in the future.

In addition, the VDH cannot explain chronological changes in MS incidence and prevalence. In the history of medicine, rickets due to vitamin D deficiency was recorded as an endemic disease in many areas such as England in the 17th century. It was a prevalent disease up to 1930, when its cause was identified, and cod liver oil and sufficient sun exposure was prescribed (Rajakumar 2003). Knowing the importance of this vitamin for health, it is routinely prescribed as a supplement for newborns and children. On the other hand, the global irradiance of ultraviolet B, which is the essential environmental factor for the production of vitamin D in the skin, has increased in past decades (Herman 2010). Therefore, if MS is actually related to vitamin D deficiency, there should be more records of the incidence and prevalence of MS, or reports of clinical manifestations that resembling MS, before 1930. We know that this is inconsistent with what has happened during the history of MS, and the fact is that MS incidence started to continuously rise since 1930 (Midgard et al. 1996).

Until recently, well designed clinical trials did not find strong direct evidence of vitamin D effects on MS incidence or the disease course (Ross et al., 2011; Kampman et al., 2012). Also, there is a notable possibility that high prevalence of vitamin D deficiency among MS patients is secondary to the disease-related disability, rather than a primary cause. More disabled patients have lower daily activity and hence less sun-exposure. Moreover, it should not be forgotten that immobilization can cause bone resorption and hypercalcemia, and therefore, lead to a secondary decrease in the production of vitamin D in the body (Sato et al. 1999). A newly published meta-analysis by Huang et al., which confirmed the lower overall bone density of MS patients in comparison to healthy controls, can be regarded as supportive evidence for this claim. They found that the reduced bone mineral density in the MS patients was associated with the disease duration and the severity of disease induced disability (Huang et al., 2014).

The VDH is unable to explain MS trends in low latitude areas. For instance, MS incidence and prevalence has significantly increased in Mexico during the past decades, in spite of the fact that Mexico is a tropical country. A recent study on this country clarified that there is no association between sunlight exposure and the risk of MS in this tropical area (Espinosa-Ramirez et al., 2014). Inability to explain the cause of the higher prevalence of MS in mountainous areas that naturally receive a higher amount of ultraviolet B radiation in comparison to the valleys is another notable failure of VDH (Risberg et al., 2011). Moreover, the VDH cannot precisely elucidate the reason for recent attenuation in the latitudinal gradient of MS prevalence, as well as phenomena such as MS epidemics (Ascherio and Munger, 2007).

Based on these facts, we believe that vitamin D may play some role in MS pathophysiology, but it seems not to be the primary factor. The effect of other potential factors, and perhaps more importantly, environmental risk factors should be evaluated.

**Geomagnetic disturbance hypothesis**

Considering the notable weaknesses of the VDH, we initially sought to investigate whether there is another environmental factor that can provide a potentially better and more comprehensive explanation about MS's epidemiological features and its pathophysiology. To attain this aim, we reviewed possible environmental factors from space to ground level and checked whether their effects on MS epidemiology had been studied or not. GMDs result from interactions of space-weather situations and the geomagnetic field (GMF). The effects of GMDs on physiological and pathophysiological issues, in comparison to other environmental factors, have generally been neglected by most biologists because the GMF and its disturbances are categorized as a very low magnetic field (VLMF) without thermal and ionizing effects (Zhadin 2001). We found that in spite of indirect evidences from some researchers in favor of a probable relationship between the GMF and MS, such as Barlow, who indicated the importance of geomagnetic coordinates, Resch, who suggested the possible role of various geophysical factors including GMDs, and the experiments of Persinger et al. (Barlow 1966; Resch 1995; Persinger, Cook, and Koren 2000), there was no comprehensive hypothesis to describe how GMDs may explain MS etiology and particularly its special epidemiological features.

Based on this evidence and geophysical characteristics of GMD, we posited the first version of our GMD hypothesis in 2012 to describe how GMDs may act as a risk factor for MS (Sajedi and Abdollahi 2012). In the following sections, we describe an improved and more complete version of the GMD hypothesis with more supportive evidence and we discuss its ability to explain certain features of MS.

It should be remembered that the GMF, even with its low magnitude, affects living beings. We know that some species sense the GMF and use it for navigation, probably by sensing the field via magnetites, i.e. ferromagnetic particles, in their CNS (Zhadin 2001). It has been shown that in a weakened GMF situation, by using a shielded chamber, hormonal disturbances occur in animal models, particularly in the blood levels of epinephrine, histamine, and serotonin (Zhadin 2001). Plants also are influenced by the GMF and a weakened GMF can cause changes in root meristems and subcellular structures like mitochondria (Zhadin 2001). It has been demonstrated that the human brain also contains magnetite (Kirschvink, Kobayashi-Kirschvink, and Woodford 1992). A growing body of evidence has revealed significant associations between myocardial infarction and brain infarcts with GMDs (Rapoport et al. 2006; Stoupel et al. 2007; Feigin et al. 2014). Some researchers believe that these effects may be induced by causing an adaptive stress reaction through the effect of GMDs on brain magnetosomes, i.e. cell membrane-bound crystals of magnetite (Kirschvink et al., 1992; Breus et al., 2008). In the same manner, we assumed that GMDs may impact important players of MS pathophysiology (Fig.1).

The CNS is regarded as the most sensitive organ to the GMF and some studies have demonstrated that CNS reactions to a magnetic field result mainly from magnetic field effects on glial cells and especially on BBB, which is a key participant in the MS pathophysiology (Zhadin, 2001). Normally, BBB acts as a highly selective barrier and strictly controls the transport of particles of blood to brain extracellular fluid. It also prohibits immune T cells from entering the brain. Due to this fact, surface antigens of CNS cells are not regarded as self-antigens by adaptive immunity and can be attacked by immune system, if immune cells enter the CNS unnecessarily, and especially if their tolerance mechanism does not work properly.

Histochemical findings about the presence of considerable iron deposits within myelin loops (LeVine and Chakrabarty, 2004) and evidence from imaging techniques about increased iron deposits around vessels in subcortical gray matters of MS patients (Lebel et al., 2012) may all be regarded as indirect clues of a probable relation between the effects of GMDs on brain magnetosomes and the pathogenesis of MS via disruption in the BBB function. However, it should be pointed out that we do not know how much of these iron deposits may be in the form of magnetite.

In 2013, Krone and Grange suggested that the higher iron in MS brains may be related to the presence of a hypothetical melanoma-like neuromelanin (MLN) (Krone and Grange, 2013). Neuromelanin is a pigment that has the ability to absorb metals, especially iron. Gathering some interesting evidence about the possible relationship between various studied risk factors of MS and neuromelanin, they proposed that MLN may be the "missing link" in the pathophysiology of MS. They indicated that changing the charge of the MLN and easing the Fenton reaction due to cosmic radiation or GMD may be the cause of the observed relationship between MS and the geomagnetic 60 degree latitude. Although some part of the latter notion needs to be revised because the level of cosmic radiation on the sea level is negligible and does not have ionizing effect, nevertheless, since neuromelanin is paramagnetic and can encompass various forms of iron, it may potentially be affected by magnetic fields and act as one of the possible mechanisms of the effect of GMDs on MS pathophysiology.

Before attacking the target, immune T cells should be activated. Activation requires two activator signals, one from the peptide/major histocompatibility complex (MHC) and, one from the co-stimulatory signal. When they exist simultaneously, they affect membrane signal transduction systems and, by starting a cascade of reactions and production of messenger molecules, eventually change gene expression and cell differentiation. Finally, adhesion and entering into the CNS are necessary for activated T cells to cause an inflammatory response and demyelination in the CNS (Kalman et al., 2008; Jandova et al., 2005).

There is evidence that the adaptive immune system can be affected by VLMF. Accordingly, a magnetic field as low as the GMF can change leukocyte behavior, activation, and adhesion by inducing the membrane-mediated signal transduction cascades, similar to when a ligand-receptor interaction activates the cell (Jandova et al., 2005; Walleczek, 1992; Simko and Mattsson, 2004; Cocek et al., 2008). The proposed mechanisms include changes of ion flux, especially $Ca^{2+}$ across a cell membrane, cyclotron resonance and dissociation of protein-ion complex by changing the quantum states of ions in their structures in the membrane proteins (Jandova et al., 2005).

Growing evidence suggests that some parts of axonal damage and the related neurodegenerative process of MS may be induced by oxidative stress due to reactive oxygen and nitrogen species generated by microglia/macrophages (Kovacic and Somanathan, 2012). Related to this fact, there is evidence that extremely low magnetic fields can enhance the release of reactive oxygen species (ROS) by T cells and macrophages (Simko and Mattsson, 2004).

Melatonin is a hormone that is secreted by the pineal gland. It has been shown that melatonin has immune-modulatory effects and it has been proposed that it may have a protective effect against MS (Sandyk, 1993). There is evidence that MS patients have a lower level of melatonin in comparison to healthy controls (Ghorbani et al., 2013), and that blood melatonin level decreases in MS patients during disease attacks (Sandyk and Awerbuch 1992). Interestingly, there is evidence that the pineal gland is very sensitive to GMD and that melatonin production is affected by GMDs (Burch, Reif, and Yost 2008; Weydahl et al. 2001). Therefore, the decrease of melatonin and the consequent alteration of self-tolerance may be another potential way that GMD provokes immune T cells to enter the CNS and attack myelin and neurons. Nevertheless, there is no direct evidence to support such a mechanism in MS. However, a series of studies by Persinger et al. on the effects of GMDs like magnetic field fluctuation on the course of EAE show that certain frequencies of GMDs can affect the behavior and infiltration of immune cells, and therefore the course and severity of this animal model of MS. They found that a 7 Hz low intensity (50 nT) magnetic field that is very similar to the background rhythm of earth magnetic field oscillation during the quiet GMF state (Schumann's resonance) has a suppressing effect on EAE.

However, drift from such situations, either in frequency or intensity, like what occurs during GMDs, can exacerbate the EAE (Persinger, Cook, and Koren, 2000; Cook, Persinger, and Koren, 2000; Kinoshameg and Persinger, 2004).

Considering these facts, GMD has the potential capability to provide the essential neuroimmunological context of MS (Fig 1). Therefore, as the core of our hypothesis, we assumed that genetically susceptible individuals would suffer from MS attacks in the geographical locations and time periods that GMD matches the sensitivity of their adaptive cell immunity, BBB, and/or pineal gland, and lasts long enough to stimulate activation, loss of tolerance and entrance of adaptive immune T cells into CNS to produce an unnecessary inflammatory response. This viewpoint can explain why the disease often has a relapsing-remitting nature.

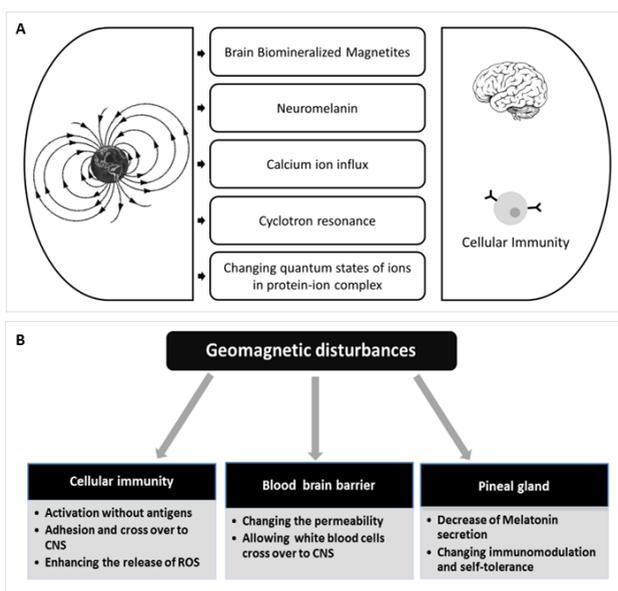

Fig. 1. Schematic description of the possible association between geomagnetic disturbance (GMD) and multiple sclerosis (MS) pathophysiology. **A**: The probable mechanisms that via them GMD may affect brain components and cellular immunity. **B**: The consequences of the effects of GMD that may contribute to MS pathophysiology. For more details please see the text.

**Explanatory ability of the GMD Hypothesis**

The GMD hypothesis has substantial ability to explain eight aspects of MS:

1) Explaining the worldwide distribution of MS

As previously mentioned, MS has a special gradient of prevalence. This epidemiological feature has been observed in various large studies and in spite of some recent reports of the attenuation of the MS prevalence gradient, any hypothesis about this disease must be able to explain this phenomenon. Accordingly, in the first step we designed a vast ecological study including 111 locations in 24 countries to test the ability of the GMD hypothesis to describe MS prevalence distribution based on the angular distance to the area with highest magnetic field disturbances. At the same time, we compared the results with the ability of the VDH that tries to describe MS prevalence by the angular distance from the equator (Sajedi and Abdollahi 2012).

The edges of auroral ovals are the areas most impacted by GMDs. Therefore, we tested whether the angular distance to the geomagnetic 60 degree latitude (AMAG60), which is the frequent location of the boundary of the auroral ovals, would describe MS prevalence or not. Our results indicated that in each continent, AMAG60 can give the best explanation for the variation of MS prevalence, better than the geographical latitude that is the basis of the VDH (Sajedi and Abdollahi, 2012). Later, a separate study by Wade et al. reported similar results (Wade, Mehta and Papitashvili, 2013).

The intensity of a GMD is related to altitude. Especially at the edge of the auroral oval, the higher the altitude, the stronger the GMD is experienced due to ionospheric disturbances. Accordingly, based on the GMD hypothesis, MS prevalence is expected to be higher in mountainous areas. Interestingly, a study in the Oppland county of Norway, which is located near the edge of the auroral oval, confirms this fact. Risberg et al. have reported that MS prevalence in inhabitants of mountain regions is higher than the rest of the county, a fact that is contradicted by the VDH (Risberg et al., 2011) but can be regarded as an important clue in favor of the GMD hypothesis.

2) The effect of time of birth

While the phenomenon is very complicated and is far beyond the scope of this article, it is well established that there are semiannual increases in GMD which take place near the time of equinoxes. In March, the earth reaches the highest southern solar latitudes, in the region that is exposed to the strong solar winds (Russell and McPherron 1973). At this time both hemispheres are affected, however, the negative interplanetary magnetic field Bx component and the positive dipole tilt of the earth cause favorable situations for the northern hemisphere high latitude reconnection phenomenon, which leads to the accentuation of transpolar arcs and magnetic field disturbances, predominantly in the auroral oval of the northern hemisphere (Kullen, Cumnock, and Karlsson, 2008). In September, the situation reverses and the reconnection phenomenon is facilitated in the southern hemisphere high latitudes (Kullen, Cumnock, and Karlsson, 2008). If we assume that GMD can increase the risk of MS by affecting adaptive cell immunity through causing memory T-cells to be activated in the future when identical temporary changes occur due to magnetic field disturbances, it could be hypothesized that genetically vulnerable individuals who were exposed to more GMDs during their CNS myelination process, when various antigens of myelin structures are exposed and are prone to be recognized by immunity, will have increased risk of developing MS in the future.

By this manner, the GMD hypothesis not only can efficiently explain the relation of the month of birth with the risk of MS, but also can describe why immigrants born in high-risk areas mostly preserve the risk of their birthplace when immigrating to low-risk locations. On the other hand, we hypothesize that genetically susceptible persons born in low-risk areas who emigrate to high-risk areas have experienced lower exposures in their fetal period that result in lower likelihood of production of memory T-cells against myelin structures, but that these individuals will exhibit increased MS risk in their new residence area due to greater frequency of exposure to GMDs, which will increase the chance of being exposed to a matched GMD with their memory T-cells sensitivity.

In 2012, Menni et al., without any consideration of our GMD hypothesis, conducted a study that evaluated the effect of solar cycles on the mortality of MS patients (Menni et al., 2012). Their original aim was to test the VDH hypothesis. However, their result showed that American white MS patients born during solar maximums have a significantly shorter lifespan than patients born in solar minimums. Surprisingly, they interpreted their result as a support of the VDH. However, they did not explain how such a result could support the VDH while, in fact, it refutes it. As we know, during solar maximums, solar irradiation, including its ultraviolet B radiation, increases along with its magnetic activity. Therefore, during solar maximums the earth receives more ultraviolet B and consequently, pregnant mothers during those periods should have a higher levels of vitamin D. According to the VDH, the risk of MS or its side effects in their infants is expected to be lower during such periods.

Conversely, Menni et al.'s finding is in favor of GMD hypothesis. Because solar maximums are the times that the earth experiences the most GMD in each solar cycle, we expect that the risk of MS and the frequency of relapses and side effects that can affect MS patients' life span would be *higher* in individuals who were born in such years. Hence, we regard their finding as positive evidence for the effect of a GMD and its relation to the time of birth of MS patients.

Recently, Janzen et al. conducted a valuable study, specifically designed to test this part of the GMD hypothesis. They evaluated the correlation of birth rates of MS patients with geomagnetic indices from 1920 to 1980 in a Canadian population that lives near the 60 degree north geomagnetic latitude. Their result confirmed a mild to moderate significant correlation between the planetary A-index of GMD and MS birth rate. Interestingly, they found a cumulative effect of exposure, that is, exposure in early life (the first three years of life) and during childhood (birth year to 12 years old) showed a higher correlation in comparison to exposure during the birth year alone (Janzen et al. 2014). Such a cumulative effect of exposure is a very strong support for the GMD hypothesis.

3) The effect of age and pregnancy

As previously mentioned, MS is very rare before the age of 15. In females, MS attacks decrease during pregnancy, but increase during post-partum. The GMD hypothesis cannot describe these features without help. When Sandyk presented the melatonin hypothesis for MS, he indicated that the lower incidence of MS in childhood and during pregnancy may be due to physiologically higher levels of melatonin secretion and its immune-modulatory effects during these periods (Sandyk, 1993, 1997). We know that GMDs decrease melatonin levels, (Burch, Reif, and Yost, 2008; Weydahl et al., 2001) and regarding the relationship between GMDs and melatonin secretion, we assume that in the period of life with a high level of melatonin (i.e. childhood and pregnancy), this decrease is not sufficient to neutralize the immune-modulatory effects of melatonin. In contrast, during physiologically lower levels of melatonin secretion (such as the post-partum state), the occurrence of GMDs causes a significant decrease of melatonin levels in susceptible individuals and enhances the loss of tolerance of their cellular immunity.

4) Historical trend of MS incidence

In the middle of the 19th century, when MS was identified by Charcot, the disease was considered a rare disease and a subject for case reports (Murray, 2005). MS incidence and hospital admission grew rapidly during the 20th century. Some researchers inferred this change as "an epidemic of recognition rather than the effect of altered biological factors" (Murray, 2005). However, from the 1930s, an increase of MS incidence was reported initially from countries like Iceland with high latitudes in the northern hemisphere that became more pronounced from 1945 to 1954 (Midgard et al., 1996). At the same time, an identical trend of changes was reported from South Africa in the southern hemisphere (Dean, 1967). Later, a remarkable increase of MS incidence and prevalence was reported from various locations such as Denmark, the Faroe Islands, Norway and Australia after 1960 (McLeod, Hammond, and Hallpike, 1994; Midgard et al., 1996). Subsequently, however, higher latitudes experienced a decreasing trend of incidence for a short period after 1965–70 (Midgard et al., 1996; Kurtzke, 2000), increasing incidence and prevalence were reported from Scotland, the United Kingdom, and the Netherlands. Such a course of events could not be easily explained by patients' survival changes or case ascertainment issues (Midgard et al., 1996). The recent birth cohort study by Ajdacic-Gross et al. confirms this increasing pattern of MS in the western countries in the first half of the 20th century (Ajdacic-Gross et al., 2013).

As we indicated earlier, VDH cannot provide an explanation for this trend. However, the historical course of MS prevalence and incidence can be explained by the GMD hypothesis. Solar activity has been observed by means of regular recording of sunspot data since 1700 (Solanki et al., 2004). Sunspot data are correlated with solar cycles, solar magnetic activity and hence with the frequency and strength of GMDs. Additionally, Solanki et al. have reconstructed sunspot data for the past 11,400 years. Their results illustrated an increasing and longstanding exceptional solar magnetic activity since 70 years ago that is unprecedented during the past 8000 years (Solanki et al., 2004). Long-term analysis of recorded GMF activities has revealed that GMDs have followed solar activity changes and have increased, a phenomenon that is known as "centennial increase of geomagnetic activity" (Mursula and Martini, 2006).

Consequently, if MS is regarded as a phenomenon related to GMDs, it not only can explain why MS or reports of clinical manifestations resembling MS were rare in the 19th century and in the medical history of all previous centuries (Murray 2005), but it also can clarify why MS incidence and prevalence have increased during the 20th century.

To test the existence of a relationship between MS incidence and GMDs, we tested the ability of the GMD hypothesis to describe the alterations of MS incidence in Tehran and western Greece (Abdollahi and Sajedi, 2014). Our evaluation showed a significant correlation between the yearly average of GMD indices and annual MS incidence alterations.

Recently, Mackenzie et al. reported the incidence and prevalence of MS in the United Kingdom from 1990 to 2010 (Mackenzie et al., 2014). We used their results and conducted a post hoc analysis of their data of MS alteration in the United Kingdom, to test whether the significant relationship between MS incidence and GMD indices is replicable or not. Interestingly, the result confirmed the existence of a similar significant association (Fig. 2) (Sajedi and Abdollahi, 2013).

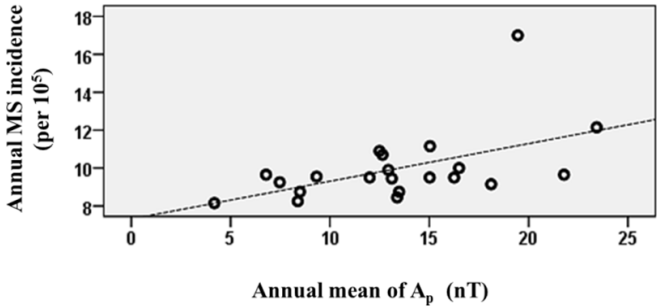

**Fig. 2**. Scatter plot of the correlation between annual averages of planetary A index ($A_P$), a geomagnetic activity index, with the annual MS incidence of the UK from 1990 to 2010 (Sajedi and Abdollahi 2013)

In spite of these studies, it is important to verify whether individual MS attacks are related to GMDs or not. The only support that we recently found in favor of this part of the GMD hypothesis, is a report by Papathanasopoulos et al. They reported a significant relationship between geomagnetic activities and hospital admissions of MS patients due to disease attacks in Greece during the 23rd solar cycle (Papathanasopoulos et al., 2016).

5) MS epidemics

Kurtzke et al. reported evidence of four MS epidemics in the small population of the Faroe Islands during 1940–1991 (Kurtzke et al., 1993). They tried to explain these epidemics by defining a hypothetical pathogen, possibly introduced by British troops in 1945 (Kurtzke et al., 1993). Nevertheless, such a pathogen has not been found (Wallin and Kurtzke, 2008).

Considering the fact that solar magnetic activities and related GMDs during the 20th solar cycle were significantly less than in the previous and subsequent cycles, and as a consequence, assuming that they were not strong enough to cause MS in susceptible Faroese, we can hypothesize that what was seen in the Faroe islands as separate MS epidemics, were reflections of solar magnetic activities and their related geomagnetic consequences on vulnerable individuals during the 17th , 18th , 19th and 21st solar cycles. Nonetheless, this statement needs to be tested by precisely superposed epoch analysis of solar magnetic activities, GMDs, and MS epidemics during the studied period.

6) The cause of significant seasonality of MS attacks

There is evidence that MS attacks significantly prone to seasonality and are more prevalent during spring and summer, particularly near the equinoxes (Meier et al., 2010). To explain this phenomenon with the GMD hypothesis, we again need the melatonin hypothesis (Sandyk, 1997) and rely on the relationship between melatonin and GMD. As mentioned above, melatonin has immune-modulatory effects. During spring and summer, the darkness is shorter and therefore, the secretion of melatonin, which is affected normally by the length of day and night, would be lesser than fall and winter. As previously mentioned, GMDs decrease melatonin secretion. Accordingly, we postulate that the further decrease of melatonin in susceptible individuals due to GMDs during spring and summer and the simultaneous immune T cell activation and entry into the CNS due to the effect of GMDs on immune cells and the BBB may be the cause of higher incidence of MS attacks during spring and summer.

7) MS and higher incidence of cardiovascular disease

Two recent studies in Sweden and Denmark have investigated the significant risk of cardiovascular diseases (CVD) in MS patients (Christiansen et al., 2010; Jadidi, Mohammadi, and Moradi, 2013). Their results showed that the risk of CVD is significantly higher in MS patients in comparison to matched controls and the general population. Previously, some studies have provided evidence about increased rates of CVDs, especially myocardial infarctions, at the time of extreme changes during or slightly after geomagnetic activities (Stoupel et al., 2007; Rapoport et al., 2006; Breus et al., 2008; Stoupel et al., 2012). Accordingly, if we regard MS patients as susceptible to GMDs, we can hypothesize that some part of the observed higher incidence of CVD in them may be related to their higher sensitivity and reaction to GMDs.

8) A Linking hypothesis

The GMD hypothesis can act as a bridge among other MS hypotheses. The melatonin hypothesis (Sandyk, 1997) and the iron accumulation hypothesis (Khalil, Teunissen, and Langkammer, 2011) can be combined with the GMD hypothesis, and by integrating their concepts, the GMD hypothesis can be transformed to the most comprehensive hypothesis about MS. The work of Krone and Grange (Krone and Grange, 2013), who proposed a comprehensive hypothesis for MS by using the GMD hypothesis and suggesting MLN as a plausible receptor of GMDs in the brain is an example of this transformational ability of the GMD hypothesis.

**GMD hypothesis limitations**

We reviewed related studies and supporting evidence to show how the GMD hypothesis may provide an explanation of MS, from its pathophysiology to the epidemiology. Nevertheless, it should be noted that these evidences are indirect or interpreted from in vitro studies. We discussed the fact that there may be a genetic basis for the cause of the sensitivity of susceptible individuals to GMDs. However, such a genetic basis, except for a new finding of the human cryptochrome CRY2 gene (Close, 2012), has not been confirmed or evaluated, and further research is needed in this area. Moreover, our supportive evidence in explaining the epidemiological features of MS by the GMD hypothesis was derived from ecological studies. It should be pointed out that such studies are often subject to the influence of ecological fallacy and that findings related to aggregate population may not always be applicable to individuals.

The GMD hypothesis has substantial unclear aspects. We do not know exactly what grade of geomagnetic disturbances is actually a risk for patients. There is evidence that very quiet states (zero activity) and geomagnetic storms can both be harmful to human health (Stoupel et al., 2012; Feigin et al., 2014). Accordingly, it seems that there is a U-shaped relationship between GMDs and human health. Such type of relationship can either be interpreted as a dose-response relationship and consequently as a positive clue for supporting the biological effects of GMDs (Calabrese and Baldwin, 2001; Calabrese, 2009) or it may be considered as a fact about which we lack sufficient knowledge of the underlying mechanisms that predispose susceptible individuals to be affected. Moreover, we do not know the exact amount of time lag between GMD occurrence and MS clinical or sub-clinical attacks.

Testing this hypothesis and clarifying its aspects would not be straightforward because: a) MS pathology is still under investigation and is not completely understood; b) the effect of race and genetic susceptibility to MS and/or to sensitivity to GMDs would influence the results of assessing the response of MS patients to GMDs; c) local factors such as crustal magnetization have significant impact on the amount of experienced GMDs in various locations, and therefore, patients in different areas may not respond similarly to a determined global index of GMDs; d) In many cases, MS demyelination occurs in silent areas of the brain and will not manifest as objective signs or symptoms. Hence, relying on patients' clinical manifestations would not be sufficient for assessing the association among GMDs and MS, and it would be necessary to monitor patients' disease activity with frequent MRI studies, which is another significant obstacle in studying the GMD hypothesis.

**Conclusion**

In this article, we described how the GMD hypothesis can explain the main features of MS. Not only does it have the ability to provide a possible explanation about MS pathophysiology, but also it has the ability to clarify the cause of its relapsing-remitting nature, historical trend, latitudinal prevalence, alterations in MS incidence, and some features such as the birth time effect and MS epidemics. It also can solve the puzzle of the longstanding failure of finding the mysterious environmental cause of MS by biochemical marker-dependent techniques and may provide the potential to make some general predictions about the disease activity, because space-weather situations and GMDs are relatively predictable.

The majority of biomedical researchers who study MS are not familiar with geomagnetic activities, and therefore, there are few studies in this area. Accordingly, it would be very valuable if medical geologists contributed to this emerging field. Clarifying the plausible link between MS and GMDs, may not only help to find new methods for preventing MS attacks, but also will likely open up new horizons for research in medical geology.

**Acknowledgements**

Some sections of this article that reviewed main abilities of the GMD hypothesis were also discussed previously in another publication of the authors (Sajedi and Abdollahi, 2012). The authors would like to indicate that they used those sections consciously, based on the fact that they are the copyright owners of the mentioned article.

The authors would like to express their gratitude to Mr. Seyed Abbas Sajedi, Mrs. Mahin Bargah Soleymani, and Mr. Kian Torab for their kind support.

**List of Abbreviations**

AMAG60: Angular distance to geomagnetic 60 degree latitude

BBB: Blood-brain-barrier

CNS: Central nervous system

CVD: Cardiovascular diseases

EAE: Experimental allergic encephalitis

GMD: Geomagnetic disturbance

GMF: Geomagnetic field

Hz: Hertz

MHC: Major histocompatibility complex

MLN: Melanoma like neuromelanin

MRI: Magnetic resonance imaging

MS: Multiple sclerosis

nT: Nano Tesla

ROS: Reactive oxygen species

VDH: Vitamin D hypothesis

VLMF: Very low magnetic field